\documentclass[aps,10pt,twocolumn,showpacs,preprintnumbers,amsmath,amssymb,prb]{revtex4-1}

\usepackage{color}
\usepackage{graphicx}
\usepackage{txfonts}

\newcommand{\C}{\mathcal{I}}
\newcommand{\expct}[2]{\left\langle #1 \right\rangle_{#2}}
\newcommand{\expcts}[2]{\langle #1 \rangle_{#2}}
\newcommand{\kmin}{k_{\rm min}}
\newcommand{\emin}{\epsilon_{\rm min}}
\newcommand{\hepsilon}{\hat{\epsilon}}
\newcommand{\ket}[1]{|#1 \rangle}
\newcommand{\bra}[1]{\langle #1 |}
\DeclareMathOperator{\sgn}{sgn}

\begin{document}

\title{Finite-temperature conductance of weakly interacting\\ quantum wires with Rashba spin-orbit coupling}

\author{Thomas~L.~Schmidt}
\affiliation{Department of Physics, University of Basel, Klingelbergstrasse 82, CH-4056 Basel, Switzerland}

\date{\today}

\begin{abstract}
We calculate the finite-temperature conductance of clean, weakly interacting one-dimensional quantum wires subject to Rashba spin-orbit coupling and a magnetic field. For chemical potentials near the center of the Zeeman gap ($\mu = 0$), two-particle scattering causes the leading deviation from the quantized conductance at finite temperatures. On the other hand, for $|\mu| > 0$, three-particle scattering processes become more relevant. These deviations are a consequence of the strongly nonlinear single-particle spectrum, and are thus not accessible using Luttinger liquid theory. We discuss the observability of these predictions in current experiments on InSb nanowires and in ``spiral liquids'', where a spontaneous ordering of the nuclear spins at low temperatures produces an effective Rashba coupling.
\end{abstract}

\pacs{73.23.-b, 71.10.Pm, 71.70.Ej, 72.10.-d}

\maketitle

\section{Introduction}

The electronic properties of one-dimensional interacting quantum wires have fascinated theorists and experimentalists for a long time. In recent years, a lot of effort has been devoted in particular to the investigation of quantum wires with strong Rashba spin-orbit coupling (SOC), mainly because these ``Rashba wires'', in the presence of a magnetic field and induced superconductivity, have been predicted to host Majorana bound states.\cite{oreg10,lutchyn10} Evidence for the latter has recently been reported in experiments.\cite{mourik12,deng12,rokhinson12}

Similar effects are also expected in a different class of materials without Rashba SOC. It was predicted several years ago that interactions between conduction electrons and nuclear spins can lead to a spontaneous magnetic ordering of the latter.\cite{braunecker09,braunecker09b} Their helical magnetic field acts back on the electrons and leads to the formation of a so-called spiral liquid with features very similar to those of a Rashba wire.\cite{braunecker12} Experimental evidence of this effect has been reported very recently using transport measurements on GaAs quantum wires.\cite{scheller13} From a theoretical point of view, spiral liquids and Rashba wires are related via a simple unitary transformation, so the results of this paper are also valid for spiral liquids.

A magnetic field lifts the spin degeneracy and causes a Zeeman shift of the single-particle spectrum. For chemical potentials inside the Zeeman gap, the transport properties of Rashba wires have been investigated using Luttinger liquid theory.\cite{gangadharaiah08,braunecker09,braunecker09b,stoudenmire11,braunecker12,meng13a} Interactions can be taken care of with bosonization, but the magnetic field and Rashba SOC produce terms which cannot be diagonalized exactly. Nevertheless, progress has been made using renormalization-group arguments, and the zero-temperature conductance has been calculated at arbitrary interaction strength.\cite{braunecker09,braunecker09b,meng13a} For a Rashba wire connected to noninteracting leads, the conductance was predicted to be quantized, $G = e^2/h$, independently of the interaction strength. This agrees with the Luttinger liquid predictions about conventional wires.\cite{ponomarenko95,safi95,maslov95}

The cornerstone of Luttinger liquid theory is the linearization of the single-particle spectrum near the Fermi points.\cite{giamarchi03} While this is an excellent approximation for calculating many thermodynamic properties at low energies, some effects such as relaxation and equilibration are missed by linearizing the spectrum.\cite{khodas07_2,barak10,karzig10,schmidt10_2,imambekov12,ristivojevic13} It was shown for conventional 1D wires that equilibration processes which change the numbers of left-moving and right-moving fermions are essential for understanding the conductance at finite temperatures.\cite{lunde07,rech09,micklitz10,micklitz11,matveev11,schuricht12} Whereas Luttinger liquid theory for a spinful system predicts a temperature-independent quantized conductance $G = 2 G_0$, where $G_0 = e^2/h$ is the conductance quantum, electron-electron interactions in the presence of a quadratic spectrum lead to a deviation $\delta G \propto -W^4 L e^{-E_F/T}$ from the quantized conductance, where $W$ is the interaction strength, $L$ the system length, $T$ the temperature, and $E_F$ the Fermi energy.\cite{lunde07} For short wires, this correction is usually small because $E_F \gg T$.

In the following, we shall calculate the conductance of a one-dimensional Rashba wire in a magnetic field using a perturbative approach in the interaction strength. For chemical potentials $\mu$ in the Zeeman gap we find that the nonlinearity of the single-particle spectrum enables equilibration processes which lead to a temperature-dependent correction to the conductance. Due to the nonparabolic form of the single-particle spectrum, the conductance correction for $\mu = 0$ is mostly due to two-particle scattering. At low temperatures it is of order $\delta G \propto -W^2 L e^{-B_z/T}$, where $B_z$ is the Zeeman energy. For $0 < |\mu| < B_z$, on the other hand, three-particle scattering provides the leading contribution, $\delta G \propto -W^4 L e^{-B_z/T} e^{|\mu|/T}$. For Zeeman energies $B_z \ll E_F$, the deviation from the quantized conductance is therefore much larger than for conventional wires without SOC and magnetic field.

It is well known that disorder can lead to a strong deviation from the quantized conductance in 1D systems at low temperatures.\cite{giamarchi03} The impact of disorder on the conductance of Rashba wires has already been investigated, in particular with regard to the effect on the observability of Majorana bound states.\cite{liu12a,bagrets12} In this paper, in contrast, we focus on wires shorter than the mean free path, where the effect of disorder can be neglected. While mean free paths in InSb Rashba wires are still of the order of $300 \text{nm}$,\cite{mourik12} much longer mean free paths of the order of $20 \mu\text{m}$ can be achieved in GaAs quantum wires, which can host spiral liquids.\cite{scheller13}

The structure of this article is a follows. In Sec.~\ref{sec:KE}, we shall introduce the necessary kinetic equation and boundary conditions, and use it for the calculation of the conductance of a noninteracting Rashba wire at finite temperature. In Sec.~\ref{sec:int}, we will use perturbation theory in the electron-electron interaction to find corrections to the conductance. In Sec.~\ref{sec:spiral}, we shall demonstrate how the results on Rashba wires carry over to system with nuclear spin order, and in Sec.~\ref{sec:conclusions}, we shall summarize our results.

\section{Kinetic equation}\label{sec:KE}

The Hamiltonian of the noninteracting Rashba wire is given by (we set $\hbar = k_B = 1$ in the following)
\begin{align}\label{eq:HRashba}
    H_0 = \sum_k \Psi^\dag_k
    \begin{pmatrix}
        \frac{k^2}{2m} - \mu - B_z & \alpha_R k \\
        \alpha_R k & \frac{k^2}{2m} -\mu + B_z
    \end{pmatrix}
    \Psi_k,
\end{align}
where $\Psi_k = \left( \psi_{\uparrow,k}, \psi_{\downarrow,k} \right)^T$ is a spinor containing spin-up and spin-down fermions. The strength of the Rashba SOC in $x$ direction is given by $\alpha_R \geq 0$, the magnetic field in $z$ direction leads to the Zeeman energy $B_z \geq 0$. For $\alpha_R = B_z = 0$, the spectra of spin-up and spin-down particles are quadratic and degenerate. A nonzero Rashba coupling $\alpha_R$ shifts the parabolas for both spin species relative to each other. The perpendicular magnetic field opens a Zeeman gap of width $2 B_z$ at $k=0$. The effects we are investigating are strongest for $|\mu| < B_z$, so we will consider chemical potentials inside the Zeeman gap in the following. Such chemical potentials have already been reached in experiments.\cite{mourik12}

\begin{figure}[t]
  \centering
  \includegraphics[width=\columnwidth]{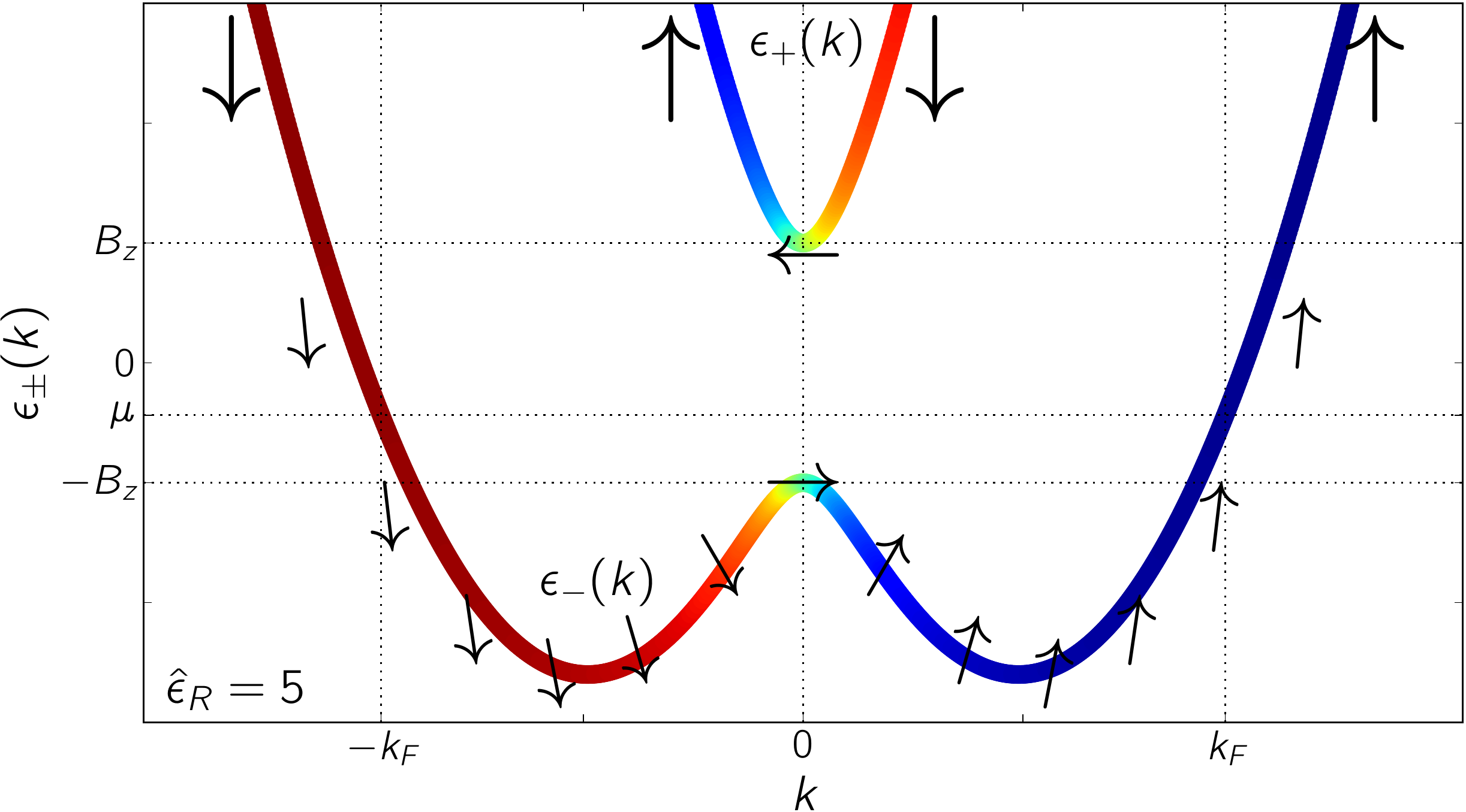}
  \caption{Single-particle spectra $\epsilon_\pm(k)$ for weak magnetic fields ($\hepsilon_R \gg 1$). The color coding and the arrows show the rotation of the spin quantization axis as a function of momentum. The chemical potential is in the Zeeman gap, $|\mu| < B_z$.}
  \label{fig:Spectrum}
\end{figure}

The Hamiltonian $H_0$ can easily be diagonalized
\begin{align}
    H_0 = \sum_k \sum_{\alpha=\pm} [\epsilon_\alpha(k) - \mu] \psi^\dag_{\alpha,k} \psi_{\alpha,k},
\end{align}
with eigenenergies and eigenstates given by, respectively,
\begin{align}\label{eq:eigenstates}
    \epsilon_\pm(k) &= \frac{k^2}{2m} \pm \sqrt{B_z^2 + \alpha_R^2 k^2}, \\
    \begin{pmatrix}
         \psi_{+,k} \\
         \psi_{-,k}
    \end{pmatrix}& =
    \begin{pmatrix}
        \sin \frac{\xi(k)}{2} & \cos \frac{\xi(k)}{2} \\
        \cos \frac{\xi(k)}{2} & -\sin \frac{\xi(k)}{2}
    \end{pmatrix}
    \begin{pmatrix}
         \psi_{\uparrow,k} \\
         \psi_{\downarrow,k}
    \end{pmatrix}     ,    \notag
\end{align}
where $\xi(k) = \arctan(\alpha_R k/B_z) \in [ -\pi/2, \pi/2]$. The relation between the energy eigenstates $\psi_{\alpha,k}$ ($\alpha = \pm$) and the spin eigenstates $\psi_{\sigma,k}$ ($\sigma = \uparrow,\downarrow$) corresponds to a rotation of the spin quantization axis with momentum. The shape of the spectrum $\epsilon_-(k)$ depends crucially on the dimensionless Rashba energy
\begin{align}
    \hepsilon_R = \frac{m\alpha_R^2}{B_z}.
\end{align}
For $\hepsilon_R > 1$, $\epsilon_-(k)$ is no longer convex. The spectrum as well as the spin orientation for the case $\hepsilon_R \gg 1$ are depicted in Fig.~\ref{fig:Spectrum}.

The zero-bias conductance of a noninteracting Rashba wire can easily be calculated using the Kubo formula. The current operator follows from the continuity equation $\partial_t \rho(x,t) + \partial_x I(x,t) = 0$, where $\rho(x) = \sum_\sigma \psi^\dag_\sigma(x) \psi_\sigma(x)$ denotes the total density,
\begin{align}
    I = -  \frac{1}{2 m i} \sum_\sigma\left( \partial_x \psi^\dag_\sigma \psi_\sigma - \psi^\dag_\sigma \partial_x \psi_\sigma    \right) + \alpha_R \sum_\sigma \psi^\dag_\sigma \psi_{-\sigma}.
\end{align}
The Kubo formula $G = \frac{i e^2}{\omega} \Pi^R(x=0,\omega)$ makes it possible to determine the conductance by calculating the retarded current-current correlation function\cite{bruus04} $\Pi^R(x,t) = -i \theta(t) \expcts{[I(x,t), I(0,0)]}{}$. At zero temperature,
\begin{align}\label{eq:cond_result}
    \frac{G(T=0)}{G_0} =\begin{cases}
        2 & \text{for } \mu > B_z, \\
        1 & \text{for } -B_z < \mu < B_z, \\
        2 & \text{for } -\emin < \mu < -B_z, \\
        0 & \text{for } \mu < -\emin,
    \end{cases}
\end{align}
where
\begin{align}
    \frac{\emin}{B_z} = \begin{cases}
        \frac{\hepsilon_R}{2} + \frac{1}{2\hepsilon_R} & \text{for } \hepsilon_R > 1,\\
        1 & \text{for } \hepsilon_R \leq 1.
    \end{cases}
\end{align}
In the regime $-B_z < \mu < B_z$, the spectrum becomes partially gapped, and the resulting conductance is reduced by a conductance quantum compared to the conductance above the gap.\cite{braunecker09,braunecker09b}

A similar calculation can be done for nonzero temperatures, but to set the stage for the discussion of interacting systems, we rederive the result using the kinetic (Boltzmann) equation. This equation is semiclassical and can be used if the mean free path is long compared to the Fermi wavelength, and the temperature exceeds the inverse lifetime of the particles.\cite{abrikosov_book} These conditions are fulfilled for clean, weakly interacting quantum wires.

In the presence of Rashba SOC and magnetic field, the single-particle states $\psi_{\alpha,k}$ diagonalize $H_0$. Therefore, we introduce the functions $f_{\alpha}(k,x)$, which denote the distribution of particles in the ``channel'' $\alpha = +,-$ with momentum $k$ at position $x$. The effect of interactions is contained in the collision integral
\begin{align}
    \C_\alpha(k,[f_+(x), f_-(x)]),
\end{align}
which determines the number of particles scattered into the state $\psi_{\alpha,k}$ per unit time, given certain distribution functions $f_\pm(k',x)$.
We consider the limit of what was called ``very short wires'' in Ref.~[\onlinecite{micklitz10}]. In this limit, the distribution functions are position-dependent because electrons do not have enough space to fully equilibrate after entering the wire from the reservoirs. We will show below that this is indeed the appropriate limit for recent experiments on Rashba wires.\cite{mourik12,deng12,rokhinson12,scheller13}

The distribution functions satisfy a coupled kinetic equation for the two channels ($\alpha = +,-$),
\begin{align}\label{eq:Boltzmann}
 v_{\alpha}(k) \partial_x f_\alpha(k,x) = \C_\alpha(k,[f_+(x), f_-(x)]),
\end{align}
where $v_{\alpha}(k) = \partial \epsilon_{\alpha}(k)/\partial k$ is the group velocity of a particle with momentum $k$ in channel $\alpha$. The presence of metallic contacts leads to boundary conditions for the distribution functions at the ends of the wire. We consider (reflectionless) adiabatic contacts,\cite{landauer87,glazman88} so right-movers (left-movers) at position $x = -L/2$ ($x = L/2$) are in thermal equilibrium with the left (right) reservoir. For $\hepsilon_R > 1$, the spectrum $\epsilon_-(k)$ has two distinct minima (see Fig.~\ref{fig:TwoParticleProcess}) and one needs to distinguish between $|k| > \kmin$ and $0 < |k| < \kmin$ for particles in the lower channel, where
\begin{align}
    \kmin =  \begin{cases}
        \frac{B_z}{\alpha_R} \sqrt{\hepsilon_R^2-1} & \text{for } \hepsilon_R > 1,\\
        0 & \text{for } \hepsilon_R \leq 1.
    \end{cases}
\end{align}
is the momentum where $\epsilon_-(k)$ reaches its minimum. The boundary conditions read
\begin{align}\label{eq:B_BC}
 f_+(k,-L/2) &= n_F[\epsilon_+(k) - \mu_+] \quad \text{for } k > 0,\notag \\
 f_-(k,-L/2) &= n_F[\epsilon_-(k) - \mu_+] \quad \text{for } k > \kmin, \notag \\
 f_-(k,-L/2) &= n_F[\epsilon_-(k) - \mu_+] \quad \text{for } -\kmin < k < 0, \notag \\
 f_+(k,+L/2) &= n_F[\epsilon_+(k) - \mu_-] \quad \text{for } k < 0,\notag \\
 f_-(k,+L/2) &= n_F[\epsilon_-(k) - \mu_-] \quad \text{for } k < -\kmin, \notag \\
 f_-(k,+L/2) &= n_F[\epsilon_-(k) - \mu_-] \quad \text{for } 0 < k < \kmin,
\end{align}
where $\mu_+$ ($\mu_-$) denotes the chemical potential of the left (right) reservoir, and $n_F(\omega) = (e^{\omega/T} + 1)^{-1}$ is the Fermi function. We use $\mu_\pm = \mu \pm eV/2$, where $V$ is the applied bias voltage. Once the distribution functions are known, the linear response current ($eV \ll T,B_z$) is obtained by
\begin{align}\label{eq:B_current}
    \expct{I(x)}{} = \frac{e}{L} \sum_{k} \sum_{\alpha = \pm} v_{\alpha}(k) f_\alpha(k,x).
\end{align}

\begin{figure}[t]
  \centering
  \includegraphics[width=\columnwidth]{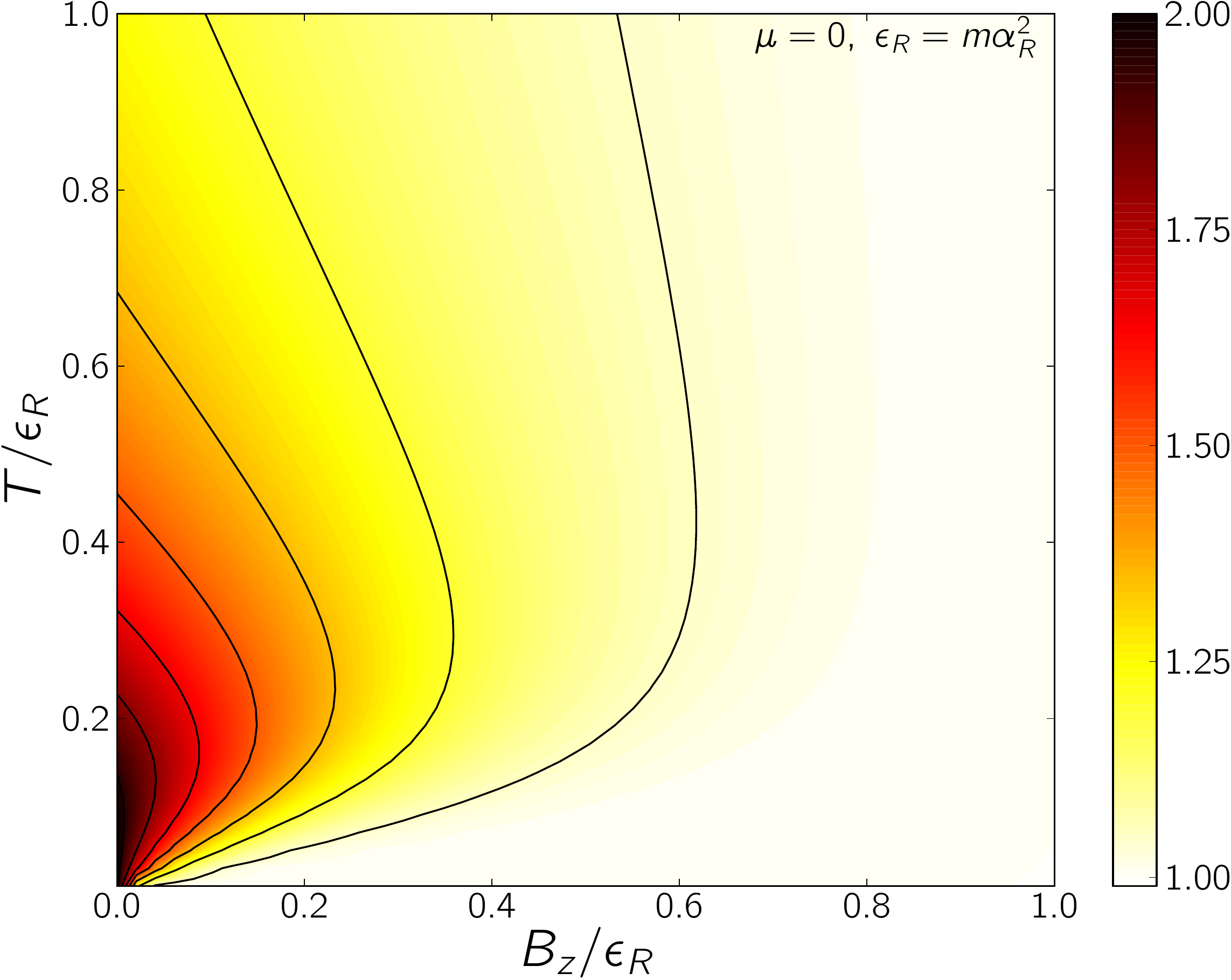}
\caption{Linear conductance of a noninteracting Rashba wire for chemical potential $\mu = 0$ as a function of temperature $T$ and magnetic field $B_z$.}
  \label{fig:ConductancePlot}
\end{figure}

Without interactions, the collision integral $\C$ vanishes. According to Eq.~(\ref{eq:Boltzmann}), the unperturbed distribution functions $f^{(0)}_\alpha(k)$ then become position-independent, and therefore coincide with their respective boundary values (\ref{eq:B_BC}) along the entire wire. Physically, this happens because for clean, noninteracting wires, the electrons retain the energy of the reservoir they originated from. Using the functions $f^{(0)}_\alpha(k)$ in Eq.~(\ref{eq:B_current}), the conductance $G = \expcts{I}{}/V$ at $V = 0$ becomes
\begin{align}\label{eq:G_noninteracting}
    \frac{G(T)}{G_0}
&=
    2 n_F(-\emin-\mu) + \sum_{\eta = \pm} \eta\ n_F(\eta B_z-\mu).
\end{align}
Taking $T\to 0$ at fixed $B_z$ and $\emin$ leads back to Eq.~(\ref{eq:cond_result}). Let us discuss briefly the case $\mu = 0$ for $\hepsilon_R > 1$. Starting from $G = G_0$ at $T=0$, a finite temperature first increases the conductance beyond $G_0$. The conductance then reaches a maximum for $T \approx B_z$ because an additional transport channel becomes available. At even higher temperatures $T \approx \emin$, the finite bandwidth becomes important and reduces the conductance again. A plot of the conductance of the noninteracting system is shown in Fig.~\ref{fig:ConductancePlot}.

\section{Interacting wires}\label{sec:int}

Next, we take into account the electron-electron interactions. We assume that the electrons interact via a density-density interaction of the form
\begin{align}
 H_{int} &= \int dx dy W(x-y) \rho(x) \rho(y).
\end{align}
Because the wire is short and interactions are weak, we can expand to lowest order in the correction to the distribution function $\delta f_\alpha(k,x) = f_\alpha(k,x) - f^{(0)}_\alpha(k)$. The boundary conditions (\ref{eq:B_BC}) are already satisfied by the unperturbed solutions $f^{(0)}_\alpha(k)$, so $\delta f_\alpha(k,x)$ vanishes at the boundaries. This allows us to express the correction to the average current due to the interactions in terms of the collision integral,
\begin{align}\label{eq:B_deltaI_final}
    \delta I
&=
    \frac{e}{2} \sum_k \sum_{\alpha=\pm}
        \zeta_\alpha(k) \C_\alpha(k,[f^{(0)}_+, f^{(0)}_-]),
\end{align}
where $\zeta_\alpha(k)$ denotes the chirality of particles with momentum $k$ in channel $\alpha$, i.e., $\zeta_+(k) = \sgn(k)$ and $\zeta_-(k) = \sgn(k) \sgn(|k| - \kmin)$. The expression for $\delta I$ has a simple physical interpretation: a particle scattered into a state with momentum $k > 0$ ($k < 0$) in the upper channel is a right-mover (left-mover) and thus contributes a positive (negative) current. Similarly, scattering a particle into a state with momentum $k$ in the lower channel gives a positive current if $k > \kmin$ or $-\kmin < k < 0$, and a negative current otherwise.

For a scattering process with $n$ incoming particles (denoted by subscripts $i1,\ldots,in$) and $n$ outgoing particles ($f1,\ldots,fn$), the collision integral reads\cite{lunde06,lunde07}
\begin{widetext}
\begin{align}\label{eq:B_Cn}
 \C^{(n)}_\alpha(k,[f^{(0)}_+, f^{(0)}_-])
&=
 - \sum_{\alpha_{i} \alpha_{f}k_{i} k_{f}} P^{(n)}_{i \to f}\delta( \epsilon_f - \epsilon_i)
     \Big[ f^{(0)}_{i1} \cdots f^{(0)}_{in}( 1 - f^{(0)}_{fn} ) \cdots ( 1 - f^{(0)}_{f1} )
- (1 - f^{(0)}_{i1}) \cdots (1 - f^{(0)}_{in}) f^{(0)}_{f1}\cdots f^{(0)}_{fn} \Big].
\end{align}
\end{widetext}
The channel indices of the incoming (outgoing) particles are denoted by $\alpha_{i(1\ldots n)}$ ($\alpha_{f(1\ldots n)}$) and their momenta are $k_{i(1\ldots n)}$ ($k_{f(1\ldots n)}$). The external channel and momentum are $\alpha \equiv \alpha_{i1}$, $k \equiv k_{i1}$, and the summation is over the remaining variables $k_i = k_{i2\ldots n}$, $k_f = k_{f1\ldots n}$, and analogously for $\alpha_i$ and $\alpha_f$. The Dirac-delta function takes care of energy conservation, with initial state and final state energies given by
\begin{align}
    \epsilon_i = \sum_{j = 1}^n \epsilon_{\alpha_{ij}}(k_{ij}),\qquad
    \epsilon_f = \sum_{j = 1}^n \epsilon_{\alpha_{fj}}(k_{fj}).
\end{align}
Moreover, $f^{(0)}_{j} \equiv f^{(0)}_{\alpha_j}(k_j)$ denotes the unperturbed distribution functions determined from Eq.~(\ref{eq:B_BC}). Finally, the transition probability between the initial and the final state follows from Fermi's golden rule,
\begin{align}\label{eq:B_defW}
 P^{(n)}_{i \to f} = 2 \pi \big| \bra{f} \hat{T} \ket{i} \big|^2
%
\end{align}
where $\hat{T} = H_{int} + H_{int} (\epsilon_i - H_0)^{-1} \hat{T}$ denotes the $T$-matrix, and $\ket{i}$ and $\ket{f}$ are the initial and final state, respectively,
\begin{align}
    \ket{i} &= \psi^\dag_{\alpha_{i1}, k_{i1}} \ldots  \psi^\dag_{\alpha_{in}, k_{in}} \ket{0}, \notag \\
    \ket{f} &= \psi^\dag_{\alpha_{f1}, k_{f1}} \ldots  \psi^\dag_{\alpha_{fn}, k_{fn}} \ket{0},
\end{align}
where $\ket{0}$ is the vacuum state. As the Hamiltonian $H_0 + H_{int}$ conserves momentum, the matrix element in Eq.~(\ref{eq:B_defW}) is nonzero only for initial and final states with the same total momentum. The bias voltage $V$ is contained in the Fermi functions in $f^{(0)}_j$. For the calculation of the linear conductance, we expand the current correction due to $n$ particle scattering to the first order in $V$, and obtain
\begin{align}\label{eq:B_deltaIn}
    \delta I^{(n)}
&=
    - \frac{\beta e^2 V}{2}  \sum_{\alpha_{i} \alpha_{f}} \sum_{k_{i1}>0} \sum_{k_{i2\ldots n}k_{f}} \sgn(k_{i1} + \alpha_{i1} \kmin)
    P^{(n)}_{i\to f}  \\
&\times \delta( \epsilon_f - \epsilon_i) F^{(n)}(\alpha_i, \alpha_f, k_i, k_f) \sum_{j=1}^n \left[
 \zeta_{\alpha_{ij}}(k_{ij}) - \zeta_{\alpha_{fj}}(k_{fj})  \right],\notag
\end{align}
and the remaining equilibrium Fermi distributions are contained in
\begin{align}\label{eq:Fn}
&    F^{(n)}(\alpha_i, \alpha_f, k_i, k_f)\\
&=
   \prod_{j=1}^n n_F[\epsilon_{\alpha_{ij}}(k_{ij})-\mu] \left\{ 1 - n_F[\epsilon_{\alpha_{fj}}(k_{fj})-\mu] \right\}. \notag
\end{align}
It follows from the sum in the second line of Eq.~(\ref{eq:B_deltaIn}) that $\delta I^{(n)} \neq 0$ only if the numbers of right-movers and left-movers, $N_R$ and $N_L$, change during a scattering process. This restricts the scattering processes which need to be considered. Even though nontrivial scattering processes involving only, say, right-movers are kinematically possible and relevant for the relaxation properties of the system,\cite{micklitz11,karzig10} they do not change the average current. Therefore, for weak interactions, it is sufficient to only consider processes with change $N_R$ and $N_L$.

Since the total momentum is conserved, scattering processes among particles near the Fermi points conserve $N_R$ and $N_L$ and thus cannot change the current. Therefore, states away from the Fermi points must be involved in the scattering. Due to the Fermi functions in Eq.~(\ref{eq:Fn}), this means that for chemical potentials $|\mu| < B_z$, the correction to the current will be exponentially suppressed as a function of temperature. In the following, we will use perturbation theory in the interaction strength to determine the current correction to the leading and next-to-leading order. In a given order of perturbation theory, we will consider those processes for which the exponential suppression is weakest.

For fermions with quadratic spectrum, energy and momentum conservation would entail that pair collisions can only lead to a permutation of the momenta of the particles. The spectra $\epsilon_\pm(k)$, on the other hand, strongly deviate from a parabolic form and thus allow particles to scatter in nontrivial ways. In particular, this means that there is a qualitative difference between the scattering probabilities for weak magnetic fields, in which case $\epsilon_-(k)$ has a local maximum at $k=0$, and for strong magnetic fields, in which case $\epsilon_\pm(k)$ start to resemble Zeeman shifted parabolas. Therefore, we will discuss these two limits separately in the following.

\section{Weak magnetic fields}

Interactions have a particularly strong effect on the conductance for weak magnetic fields, i.e., $\hepsilon_R = m\alpha_R^2/B_z \gg 1$, because the lower channel $\epsilon_-(k)$ develops a local maximum at $k=0$.

The position of the Fermi energy $\mu$ is crucial for determining the kinematically allowed scattering processes. For $\mu \approx 0$, two-particle scattering is possible and yields a conductance correction $\delta G \propto -W^2 L e^{-B_z/T} e^{-|\mu|/T}$ already to the second order in the interaction amplitude. These processes are strongest at $\mu = 0$, and become suppressed for $|\mu| > 0$. In contrast, three-particle scattering yields a correction $\delta G \propto -W^4 L e^{-B_z/T} e^{|\mu|/T}$. Because of this weaker exponential suppression, three-particle processes therefore become the most relevant scattering mechanism for $|\mu| \to B_z$. In the following, we will describe the conductance corrections due to two-particle and three-particle scattering processes for weak magnetic fields, $\hepsilon_R \gg 1$.

\subsection{Two-particle scattering}

\begin{figure}[t]
  \centering
  \includegraphics[width=\columnwidth]{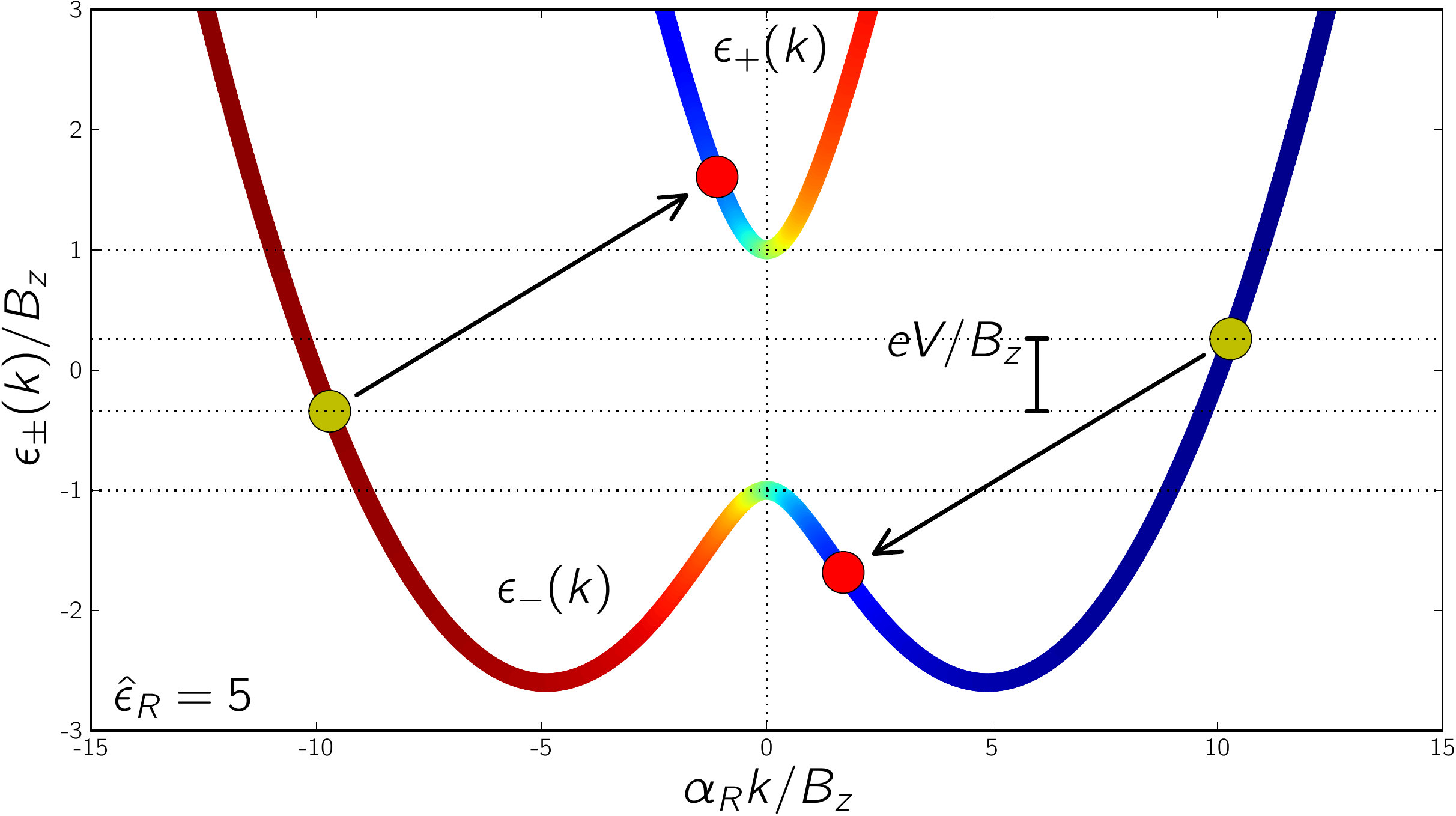}
  \caption{Two-particle scattering process for $\mu \approx 0$. This process contributes to the conductance correction because it changes the numbers of right and left-movers.}
  \label{fig:TwoParticleProcess}
\end{figure}

To the leading order in the interaction strength, we use $\hat{T} = H_{int}$ in the scattering probability (\ref{eq:B_defW}). This corresponds to considering a single scattering event, so the initial and final states $\ket{i}$ and $\ket{f}$ each contain two particles. Let us consider the consequences of energy and momentum conservation for this type of scattering. The most relevant processes at low temperatures involves initial or final states with particles near the Fermi points $k \approx \pm k_F$. Nontrivial scattering processes are possible, e.g., if one of the initial particles is near the right Fermi point and the other one near the left Fermi point, such that the total initial momentum is close to zero. In that case, these two particles can scatter into two particles with momenta close to $k=0$, one in the upper channel and one in the lower channel. The latter two can have the same chirality, so scattering can change $N_R$ and $N_L$. An example of such a process is depicted in Fig.~\ref{fig:TwoParticleProcess}.

Obviously, the bottleneck for this process is the existence of an empty state near $k=0$, i.e., below the Fermi energy. The probability of finding such a state at $k=0$ for $\mu = 0$ is proportional to $e^{-B_z/T}$. As a consequence, the conductance correction at low temperatures is given by (for $\mu = 0$, $\hepsilon_R \gg 1$, and $T < B_z$)
\begin{align}\label{eq:deltaG}
    \frac{\delta G(\mu = 0)}{G_0}
&\approx
    - n_0 \left(\frac{W(k_F)}{\alpha_R}\right)^2 \left(\frac{L B_z}{\alpha_R}\right) \sqrt{\frac{T}{B_z}} e^{-B_z/T},
\end{align}
where $n_0 \approx 3$ is a numerical prefactor which arises from an integral over a product of Fermi functions, see Eq.~(\ref{eq:B_Cn}).

To understand the nontrivial temperature dependence of Eq.~(\ref{eq:deltaG}) physically, we consider a simplified model where we linearize the spectrum near the two Fermi points $\pm k_F$, and approximate it as parabolic near $k = 0$,
\begin{align}\label{eq:spec_linear}
    \epsilon_-(k) &= v_F (\pm k - k_F)  & (\text{for } k \approx \pm k_F), \notag \\
    \epsilon_\pm(k) &= \frac{k^2}{2m_\pm} \pm B_z  & (\text{for } k \approx 0),
\end{align}
where $v_F =\partial_k \epsilon_-(k)|_{k = k_F}$ is the Fermi velocity. Moreover, in the limit $\hepsilon_R \gg 1$, we can use $m_+ \approx - m_- \equiv m^*$, where $m^* = m/\hepsilon_R > 0$ is an effective band mass which is much smaller than the mass $m$ of the physical fermions. Let us denote the momenta of the initial state particles in the lower channel near the Fermi points by $k_{i1} = k_F + p_{i1}$ and $k_{i2} = -k_F + p_{i2}$. The final state particle $f1$ ($f2$) is in the lower (upper) channel and has momentum $p_{f1}$ ($p_{f2}$), where $|p_{(i,f)(1,2)}| \ll k_F$. For given initial state momenta, energy and momentum conservation allow a unique final state momentum,
\begin{align}\label{eq:Rpf1}
    p_{f1} = p - \frac{m^* v_F q}{2 p},
\end{align}
where $p = (p_{i1} + p_{i2})/2$ and $q = p_{i1} - p_{i2}$ denote center-of-mass and relative momentum, respectively, of the initial state particles. To generate a current correction, both final state particles must have the same chirality. The corresponding condition $p_{f1} p_{f2} < 0$ translates to
\begin{align}\label{eq:Rplimit}
    p^2 < \left| \frac{m^* v_F q}{2} \right|.
\end{align}
The bottleneck for this process is the generation of the final state particle $f1$ deep in the Fermi sea. Due to the Fermi function, this probability is suppressed as $\exp\{ - [ p_{f1}^2/(2m^*) + B_z ]/T \}$ at low temperatures. This makes it favorable to create the particle at $p_{f1} = 0$. According to Eq.~(\ref{eq:Rpf1}), this corresponds to the upper limit allowed by Eq.~(\ref{eq:Rplimit})

The temperature dependence of Eq.~(\ref{eq:deltaG}) can now be understood as follows: the total scattering probability of a given incoming particle involves three integrations over the momenta of the three other particles. Two of these integrals are cancelled by energy and momentum conservation, leaving one integration over a momentum range of width $\propto T/v_F$. The energetically most favorable process involves creating a particle at $p_{f1} \approx 0$ with energy $\epsilon_{f1} \lessapprox -B_z$. Due to the van-Hove singularity in the density of states at energy $-B_z$, the probability for finding an available state at energy $\epsilon_{f1}$ is given by $e^{\epsilon_{f1}/T}/\sqrt{|\epsilon_{f1}+B_z|}$. The integration over a small range of energies $0 < \epsilon_{f1} + B_z < T$ thus yields the $\sqrt{T} e^{-B_z/T}$ in the prefactor of Eq.~(\ref{eq:deltaG}).

For $\mu \neq 0$ the two-particle scattering processes illustrated in Fig.~\ref{fig:TwoParticleProcess} are suppressed: for $\mu < 0$, the energy of the initial state is insufficient to create a final state particle in the lower channel at $p_{f1} = 0$ and a final state particle in the upper band. On the other hand, for $\mu > 0$, the energy would be sufficient, but the final state particle in the lower channel at $p_{f1} = 0$ now lies deeper in the Fermi sea. In either case, this leads to an additional exponential suppression $\delta G \propto -e^{-B_z/T} e^{-|\mu|/T}$. In this limit $|\mu| \lessapprox B_z$, it turns out that three-particle scattering may contribute a stronger correction to the conductance.

\subsection{Three-particle scattering}

\begin{figure}[t]
  \centering
  \includegraphics[width=\columnwidth]{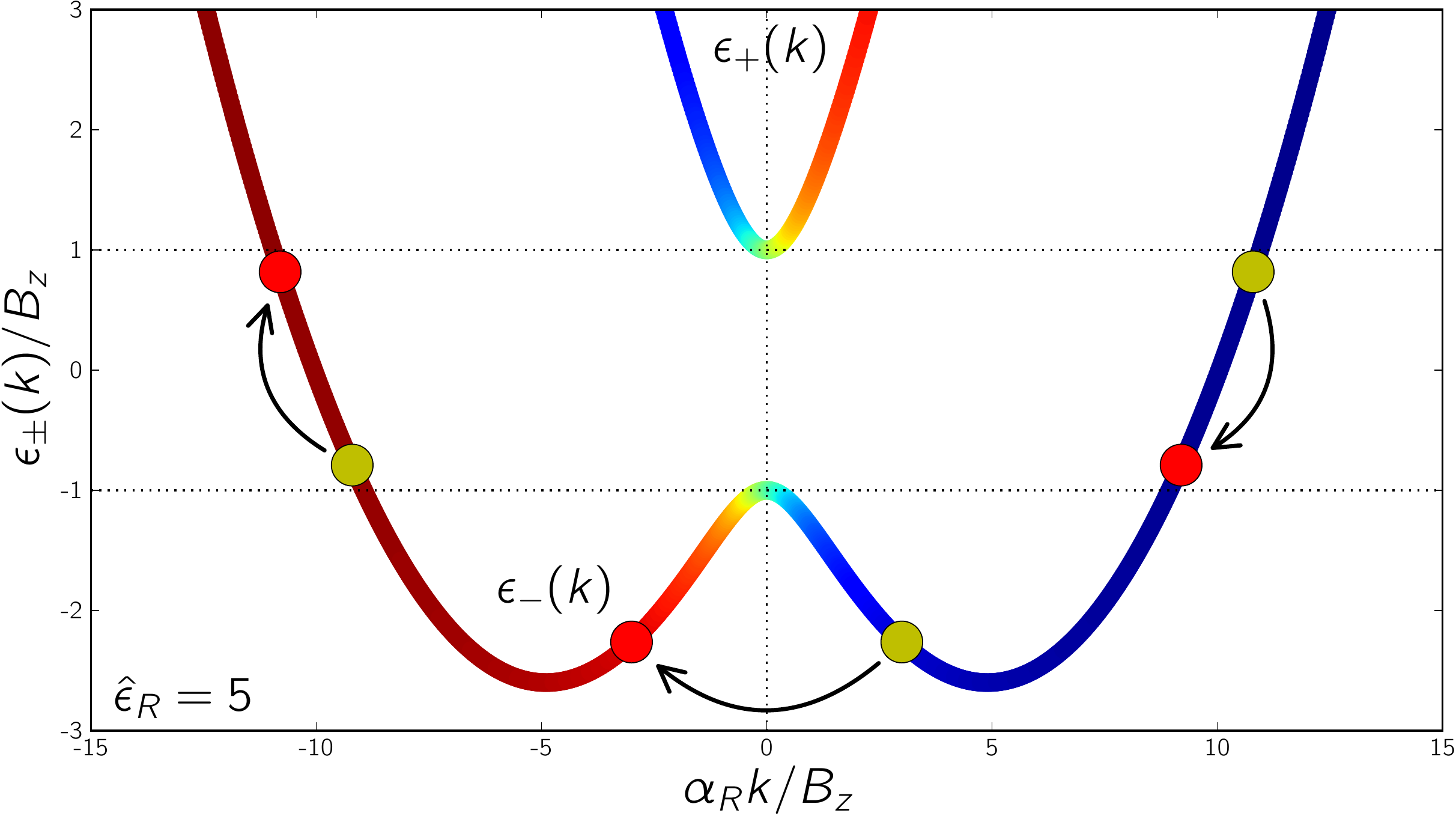}
  \caption{Three-particle scattering process for $|\mu| < B_z$. This process contributes to the conductance correction because it changes the numbers of right and left-movers.}
  \label{fig:ThreeParticleProcess}
\end{figure}

Three-particle scattering becomes the leading contribution for $|\mu|\lessapprox B_z$. A possible process is shown in Fig.~\ref{fig:ThreeParticleProcess}. It starts with an initial state containing two particles at opposite Fermi points, and one particle at $k \approx 0$. The final state still contains two particles at opposite Fermi points, but the particle near $k\approx 0$ has changed direction. Thus, this process contributes to the current correction.

The collision integral for three-particle scattering is analogous to Eq.~(\ref{eq:B_Cn}), but it requires taking into account the $T$-matrix expansion up to the second order in $H_{int}$. Using Wick's theorem for the transition probabilities $P^{(3)}_{i \to f}$ yields a large number of terms which encompass all possible direct and exchange diagrams. After identifying and calculating all contributing diagrams, the collision integral (\ref{eq:B_Cn}) yields the current correction via Eqs.~(\ref{eq:Boltzmann}) and (\ref{eq:B_current}). Due to the large number of second-order diagrams involved, the analytic result becomes very lengthy and depends on the details of the interaction potential $W(k)$. In order to simplify the calculation, we consider two particular interaction potentials.

First, let us consider the case of a pointlike interaction potential, $W(k) = W$. Since Rashba spin-orbit coupling destroys the integrability of the system, scattering can even arise for this type of potential. As a consequence, one obtains (for $|\mu| \lessapprox B_z$, $\hepsilon_R \gg 1$, and $T < B_z - |\mu|$)
\begin{align}\label{eq:deltaG3}
    \frac{\delta G}{G_0} &= - n_1 \left(\frac{W}{\alpha_R} \right)^4 \left(\frac{L B_z}{\alpha_R} \right) \left(\frac{T}{B_z}\right)^3 e^{-B_z/T} e^{|\mu|/T},
\end{align}
where $n_1 \approx 35$. The pointlike interaction potential allows scattering processes for which the matrix elements remain finite even at zero energy. Therefore, the temperature dependence can be understood as follows: the scattering probability for an incoming particle involves five integrations over the momenta of the remaining particles. Two of these are cancelled due to energy and momentum conservation, and the three remaining ones produce the factor $T^3$.

Moreover, let us consider the case of a long-ranged potential where $W(k_F) \ll W(0)$, such that $W(k_F)$ is negligible. This condition can be fulfilled for a screened Coulomb potential. To the fourth order in the interaction strength, one then finds (for $|\mu| \lessapprox B_z$, $\hepsilon_R \gg 1$, and $T < B_z - |\mu|$),
\begin{align}\label{eq:deltaG4}
    \frac{\delta G}{G_0} &= -n_1' \left(\frac{W(0)}{\alpha_R}\right)^4  \left(\frac{L B_z}{\alpha_R}\right) \left(\frac{T}{B_z}\right)^7 e^{-B_z/T}e^{|\mu|/T},
\end{align}
where $n'_1 \approx 10^3$. This result is more strongly suppressed for low temperatures than Eq.~(\ref{eq:deltaG3}) because scattering processes which involve a momentum exchange $k_F$ are no longer possible. The strongest scattering process at low temperatures involves an intermediate state where the initial state particle at $k \approx 0$ is scattered into a virtual state at momentum $k' \approx 0$ in the opposite channel. Due to the spin structure of the eigenstates, see Eq.~(\ref{eq:eigenstates}), the corresponding amplitude has a prefactor
\begin{align}
    \left[\sin\left(\frac{\xi(k) - \xi(k')}{2}\right) \right]^2 \propto \left[\frac{\alpha_R}{2 B_z} (k - k')\right]^2
\end{align}
for $|k|, |k'| \ll B_z/\alpha_R$. Therefore, the scattering probability contributes another factor $\propto T^4$ compared to Eq.~(\ref{eq:deltaG3}). This explains the prefactor $\propto T^7$ in Eq.~(\ref{eq:deltaG4}).

\subsection{Experimental visibility in Rashba wires}

Let us briefly assess the experimental visibility of the correction~(\ref{eq:deltaG}) in a Rashba wire using the parameters of Ref.~[\onlinecite{mourik12}]. We consider a clean InSb wire with length $L \approx 2 \mu$m, Rashba SOC $\alpha_R = 0.2 \text{ eV\AA}$ and Zeeman energy $B_z = 1.5 B \text{ meV/T}$, where $B$ is the magnetic field. In the limit of weak magnetic fields and for $\mu \approx 0$, we can assume $v_F \approx \alpha_R$ for the Fermi velocity.\cite{rainis12} A Luttinger parameter $K \approx 0.9$, which corresponds to weak interactions, then leads to $W(k_F) \approx v_F/2$. Assuming a temperature $T \approx 50 \text{ mK}$, the correction $\delta G/G_0 \approx -0.03$ for magnetic fields $B \approx 5 \text{ mT}$ (such that $B_z/T \approx 1.7$). Therefore, for Zeeman energies $B_z \gtrapprox T$ the correction $\delta G$ is significant. Moreover, due to its length dependence, the interaction contribution (\ref{eq:deltaG}) can be experimentally distinguished from the (length-independent) noninteracting correction (\ref{eq:G_noninteracting}) by comparing wires of different lengths. In addition, adapting the estimates of Ref.~[\onlinecite{micklitz10}] and using the same parameters, one finds that the maximum length up to which our model of ``very short wires'' applies is $l_0 \approx 8 \mu$m, so this limit is indeed appropriate for current experiments.

\section{Strong magnetic fields}

Finally, let us consider briefly the limit of strong magnetic fields, $\hepsilon_R = m\alpha_R^2/B_z \ll 1$. In this case, Zeeman splitting dominates over the Rashba SOC, so the spectrum becomes increasingly parabolic. In particular, the local maximum of $\epsilon_-(k)$ at $k=0$ turns into a global minimum and $\epsilon_\pm(k)$ become convex functions of momentum.
The leading process at low temperature is again brought about by scattering two particles near $\pm k_F$ from the lower channel into a final state which contains one particle in the upper channel, and one particle in the lower channel near $k = 0$.

Since one of the scattered particles must flip its spin, this correction vanishes for $\alpha_R = 0$. To leading order in $\alpha_R$, one finds (for $\mu = 0$, $\hepsilon_R \ll 1$, and $T < B_z$)
\begin{align}\label{eq:deltaG_strong}
     \frac{\delta G(\mu = 0)}{G_0}
&\approx
   - n_2 \left( \frac{m \alpha_R^2}{B_z} \right) \left( \frac{m W(k_F)^2}{B_z} \right)
     \sqrt{ m B_z L^2} \left( \frac{T}{B_z} \right)  e^{-B_z/T},
\end{align}
where $n_2 \approx 1.25$. The correction due to two-particle scattering again decreases exponentially for $\mu \neq 0$, $\delta G \propto e^{-B_z/T} e^{-|\mu|/T}$.

The temperature dependence of Eq.~(\ref{eq:deltaG_strong}) can be understood physically as follows. Similarly to  Eq.~(\ref{eq:spec_linear}), we first linearize the spectrum near $|k| \approx k_F$ and approximate it to quadratic order near $k \approx 0$, but in this case $m_+ \approx m_- \approx m$. In terms of the center-of-mass momentum $p$ and the relative momentum $q$ of the initial state particles, energy and momentum conservation admit two possible momenta for the final state particle in the lower band,
\begin{align}
    p_{f1,\pm} = p \pm \sqrt{m v_F q - p^2}.
\end{align}
Scattering is thus possible if $q > 0$ and $p^2 < m v_F q$. Moreover, the condition that the two particles in the final state have the same chirality ($p_{f1} p_{f2} > 0$) now translates to $p > \sqrt{m v_F q/2}$. As a consequence, scattering which changes the current is possible in the range
\begin{align}
    \sqrt{\frac{m v_F q}{2}} < p < \sqrt{m v_F q}.
\end{align}
The probability for finding an empty state in the lower channel near $k \approx 0$ is proportional to $e^{\epsilon_{f1}/T}$, where $\epsilon_{f1} = p^2_{f1}/(2m) - B_z <  0$. This makes it favorable to create the final state particle in the lower channel at the highest allowed momentum, i.e., $p_{f1} = \sqrt{2 m v_F q}$. Note that in contrast to the discussion after Eq.~(\ref{eq:Rplimit}), there is no van-Hove singularity at this momentum. Hence, a missing factor $1/\sqrt{T}$ compared to Eq.~(\ref{eq:deltaG}) indeed gives a conductance proportional to $T e^{-B_z/T}$.

For $|\mu| \lessapprox B_z$ and $\hepsilon_R \ll 1$, three-particle scattering yields again the most important correction. For $\alpha_R = 0$ and pointlike interaction potential $W(k) = \text{const.}$, the system becomes integrable,\cite{yang67,gaudin67,sutherland04} and we find that the backscattering amplitude and thus the current correction vanishes. On the other hand, for a generic finite-range interaction, a nonzero contribution arises. For $\mu \approx B_z$ ($\mu \approx -B_z$), the most relevant process involves an initial state with two particles near the Fermi points and one particle in the upper (lower) channel near $k = 0$. Scattering with small momentum transfer changes the direction of the particle at $k = 0$. For $\alpha_R = 0$, this process is identical to the one considered in Ref.~[\onlinecite{lunde07}]. One thus finds a correction $\delta G \propto -e^{-B_z/T} e^{|\mu|/T}$, with a prefactor that depends on the detailed form of the interaction potential $W(k)$.\cite{lunde07}

\section{Systems with nuclear spin order}\label{sec:spiral}

It was predicted several years ago that systems with nuclear spins and hyperfine interaction, even in the absence of Rashba spin-orbit coupling, can have physical properties which are very similar to those of Rashba wires. This makes it possible to realize the Hamiltonian~(\ref{eq:HRashba}), e.g., in conventional GaAs systems.\cite{braunecker09,braunecker09b} In these systems, an effective Rashba spin-orbit coupling is produced by the interplay between the nuclear spins and the conduction electrons, and the corresponding state has been called a ``spiral Luttinger liquid''.\cite{braunecker12} Indeed, the conduction electrons allow distant nuclear spins to interact via the RKKY interaction. At low temperatures, this allows the nuclear spins to order in a helical arrangement. The resulting helical magnetic fields acts back on the conduction electrons, and has a similar effect as Rashba spin-orbit coupling. Signatures of such a helical nuclear spin ordering have recently been observed in experiments.\cite{scheller13}

The starting point is a spin-degenerate Hamiltonian for the conduction band electrons in a quantum wire,
\begin{align}
    H_0 &= \sum_{\sigma} \int dx \psi^\dag_\sigma(x) \left( - \frac{\partial_x^2}{2m} - \mu \right) \psi_\sigma(x),
\end{align}
with chemical potential $\mu = k_F^2/(2m)$. The polarization of the nuclear spins creates a helical magnetic field $\vec{B}(x) = B \left[ \cos(2 k_F x) \vec{e}_x + \sin(2 k_F x) \vec{e}_y \right]$ which rotates with a wave vector $2k_F$ in the spin x-y plane. This corresponds to the Hamiltonian,\cite{braunecker09,braunecker09b}
\begin{align}
    H_B &= \sum_{\sigma\sigma'} \int dx \psi^\dag_\sigma(x) \left[ \vec{B}(x) \cdot \vec{S} \right]_{\sigma\sigma'} \psi_{\sigma'}(x),
\end{align}
where $\vec{S}$ is the vector of Pauli matrices. The strength of the magnetic field $B$ depends of the hyperfine interaction and the magnetization of the nuclear spins, which is in turn temperature-dependent. The Hamiltonian $H_B$ allows scattering of spin-down particles with momentum near $k_F$ into spin-up particles with momentum $-k_F$, and vice versa, and thus opens a partial gap at the Fermi points.

The unitary transformation $U = e^{i A}$ with $A = k_F \sum_{\sigma} \sigma \int dx x \psi^\dag_\sigma \psi_{\sigma}$ maps the Hamiltonian $H_0 + H_B$ onto a Rashba Hamiltonian. A subsequent spin axis rotation with the unitary transformation $W = \sigma_x e^{i \pi \sigma_y/4}$ makes the resulting Hamiltonian identical to Eq.~(\ref{eq:HRashba}) with the parameters,
\begin{align}\label{eq:para_spiral}
    \alpha_R = \frac{k_F}{m} = v_F, \qquad
    B_z = B, \qquad
    \mu = 0,
\end{align}
where $v_F$ is the Fermi velocity. The mapping between spiral liquids and Rashba systems works even in the presence of density-density interactions, because the unitary transformation $U$ commutes with the spin densities $\psi^\dag_{\sigma}(x) \psi_{\sigma}(x)$.

As the helical magnetic field is much smaller than the Fermi energy, $\hepsilon_R \gg 1$ is the experimentally relevant regime for spiral liquids. Therefore, because $\mu = 0$, we expect the temperature-dependent conductance $G(T)$ of a clean, weakly interacting spiral liquid at temperatures below the ordering temperature of the nuclear spins to be described by Eq.~(\ref{eq:deltaG}) with the parameters~(\ref{eq:para_spiral}).

\section{Conclusions}\label{sec:conclusions}

In conclusion, we have calculated the temperature-dependent conductance of a clean, weakly interacting quantum wire subject to Rashba spin-orbit coupling and a perpendicular magnetic field. We found that at nonzero temperatures, interactions cause length-dependent corrections $\delta G(T)$ to the quantized conductance, which are not captured by Luttinger liquid theory because they rely on the nonlinearity of the spectrum. For chemical potential $\mu = 0$, two-particle scattering is the most important process. Three-particle processes become increasingly relevant for $|\mu| \lessapprox B_z$. Using realistic experimental parameters, we estimated that for $\mu = 0$ the correction $\delta G$ should be experimentally observable for small Zeeman energies $B_z \gtrapprox T$.

\acknowledgments
The author acknowledges helpful discussions with T.~Meng, D.~Rainis, D.~Loss, and L.~Glazman. This work was financially supported by the Swiss NSF.

\bibliography{paperShortWires}

\begin{thebibliography}{39}%
\makeatletter
\providecommand \@ifxundefined [1]{%
 \@ifx{#1\undefined}
}%
\providecommand \@ifnum [1]{%
 \ifnum #1\expandafter \@firstoftwo
 \else \expandafter \@secondoftwo
 \fi
}%
\providecommand \@ifx [1]{%
 \ifx #1\expandafter \@firstoftwo
 \else \expandafter \@secondoftwo
 \fi
}%
\providecommand \natexlab [1]{#1}%
\providecommand \enquote  [1]{``#1''}%
\providecommand \bibnamefont  [1]{#1}%
\providecommand \bibfnamefont [1]{#1}%
\providecommand \citenamefont [1]{#1}%
\providecommand \href@noop [0]{\@secondoftwo}%
\providecommand \href [0]{\begingroup \@sanitize@url \@href}%
\providecommand \@href[1]{\@@startlink{#1}\@@href}%
\providecommand \@@href[1]{\endgroup#1\@@endlink}%
\providecommand \@sanitize@url [0]{\catcode `\\12\catcode `\$12\catcode
  `\&12\catcode `\#12\catcode `\^12\catcode `\_12\catcode `\%12\relax}%
\providecommand \@@startlink[1]{}%
\providecommand \@@endlink[0]{}%
\providecommand \url  [0]{\begingroup\@sanitize@url \@url }%
\providecommand \@url [1]{\endgroup\@href {#1}{\urlprefix }}%
\providecommand \urlprefix  [0]{URL }%
\providecommand \Eprint [0]{\href }%
\providecommand \doibase [0]{http://dx.doi.org/}%
\providecommand \selectlanguage [0]{\@gobble}%
\providecommand \bibinfo  [0]{\@secondoftwo}%
\providecommand \bibfield  [0]{\@secondoftwo}%
\providecommand \translation [1]{[#1]}%
\providecommand \BibitemOpen [0]{}%
\providecommand \bibitemStop [0]{}%
\providecommand \bibitemNoStop [0]{.\EOS\space}%
\providecommand \EOS [0]{\spacefactor3000\relax}%
\providecommand \BibitemShut  [1]{\csname bibitem#1\endcsname}%
\let\auto@bib@innerbib\@empty
\bibitem [{\citenamefont {Oreg}\ \emph {et~al.}(2010)\citenamefont {Oreg},
  \citenamefont {Refael},\ and\ \citenamefont {von Oppen}}]{oreg10}%
  \BibitemOpen
  \bibfield  {author} {\bibinfo {author} {\bibfnamefont {Y.}~\bibnamefont
  {Oreg}}, \bibinfo {author} {\bibfnamefont {G.}~\bibnamefont {Refael}}, \ and\
  \bibinfo {author} {\bibfnamefont {F.}~\bibnamefont {von Oppen}},\ }\href
  {\doibase 10.1103/PhysRevLett.105.177002} {\bibfield  {journal} {\bibinfo
  {journal} {Phys. Rev. Lett.}\ }\textbf {\bibinfo {volume} {105}},\ \bibinfo
  {pages} {177002} (\bibinfo {year} {2010})}\BibitemShut {NoStop}%
\bibitem [{\citenamefont {Lutchyn}\ \emph {et~al.}(2010)\citenamefont
  {Lutchyn}, \citenamefont {Sau},\ and\ \citenamefont {Das~Sarma}}]{lutchyn10}%
  \BibitemOpen
  \bibfield  {author} {\bibinfo {author} {\bibfnamefont {R.~M.}\ \bibnamefont
  {Lutchyn}}, \bibinfo {author} {\bibfnamefont {J.~D.}\ \bibnamefont {Sau}}, \
  and\ \bibinfo {author} {\bibfnamefont {S.}~\bibnamefont {Das~Sarma}},\ }\href
  {\doibase 10.1103/PhysRevLett.105.077001} {\bibfield  {journal} {\bibinfo
  {journal} {Phys. Rev. Lett.}\ }\textbf {\bibinfo {volume} {105}},\ \bibinfo
  {pages} {077001} (\bibinfo {year} {2010})}\BibitemShut {NoStop}%
\bibitem [{\citenamefont {Mourik}\ \emph {et~al.}(2012)\citenamefont {Mourik},
  \citenamefont {Zuo}, \citenamefont {Frolov}, \citenamefont {Plissard},
  \citenamefont {Bakkers},\ and\ \citenamefont {Kouwenhoven}}]{mourik12}%
  \BibitemOpen
  \bibfield  {author} {\bibinfo {author} {\bibfnamefont {V.}~\bibnamefont
  {Mourik}}, \bibinfo {author} {\bibfnamefont {K.}~\bibnamefont {Zuo}},
  \bibinfo {author} {\bibfnamefont {S.~M.}\ \bibnamefont {Frolov}}, \bibinfo
  {author} {\bibfnamefont {S.~R.}\ \bibnamefont {Plissard}}, \bibinfo {author}
  {\bibfnamefont {E.~P. A.~M.}\ \bibnamefont {Bakkers}}, \ and\ \bibinfo
  {author} {\bibfnamefont {L.~P.}\ \bibnamefont {Kouwenhoven}},\ }\href
  {\doibase 10.1126/science.1222360} {\bibfield  {journal} {\bibinfo  {journal}
  {Science}\ }\textbf {\bibinfo {volume} {336}},\ \bibinfo {pages} {1003}
  (\bibinfo {year} {2012})}\BibitemShut {NoStop}%
\bibitem [{\citenamefont {Deng}\ \emph {et~al.}(2012)\citenamefont {Deng},
  \citenamefont {Yu}, \citenamefont {Huang}, \citenamefont {Larsson},
  \citenamefont {Caroff},\ and\ \citenamefont {Xu}}]{deng12}%
  \BibitemOpen
  \bibfield  {author} {\bibinfo {author} {\bibfnamefont {M.~T.}\ \bibnamefont
  {Deng}}, \bibinfo {author} {\bibfnamefont {C.~L.}\ \bibnamefont {Yu}},
  \bibinfo {author} {\bibfnamefont {G.~Y.}\ \bibnamefont {Huang}}, \bibinfo
  {author} {\bibfnamefont {M.}~\bibnamefont {Larsson}}, \bibinfo {author}
  {\bibfnamefont {P.}~\bibnamefont {Caroff}}, \ and\ \bibinfo {author}
  {\bibfnamefont {H.~Q.}\ \bibnamefont {Xu}},\ }\href {\doibase
  10.1021/nl303758w} {\bibfield  {journal} {\bibinfo  {journal} {Nano Letters}\
  }\textbf {\bibinfo {volume} {12}},\ \bibinfo {pages} {6414} (\bibinfo {year}
  {2012})}\BibitemShut {NoStop}%
\bibitem [{\citenamefont {Rokhinson}\ \emph {et~al.}(2012)\citenamefont
  {Rokhinson}, \citenamefont {Liu},\ and\ \citenamefont
  {Furdyna}}]{rokhinson12}%
  \BibitemOpen
  \bibfield  {author} {\bibinfo {author} {\bibfnamefont {L.~P.}\ \bibnamefont
  {Rokhinson}}, \bibinfo {author} {\bibfnamefont {X.}~\bibnamefont {Liu}}, \
  and\ \bibinfo {author} {\bibfnamefont {J.~K.}\ \bibnamefont {Furdyna}},\
  }\href {\doibase 10.1038/nphys2429} {\bibfield  {journal} {\bibinfo
  {journal} {Nat. Phys.}\ }\textbf {\bibinfo {volume} {8}},\ \bibinfo {pages}
  {795} (\bibinfo {year} {2012})}\BibitemShut {NoStop}%
\bibitem [{\citenamefont {Braunecker}\ \emph
  {et~al.}(2009{\natexlab{a}})\citenamefont {Braunecker}, \citenamefont
  {Simon},\ and\ \citenamefont {Loss}}]{braunecker09}%
  \BibitemOpen
  \bibfield  {author} {\bibinfo {author} {\bibfnamefont {B.}~\bibnamefont
  {Braunecker}}, \bibinfo {author} {\bibfnamefont {P.}~\bibnamefont {Simon}}, \
  and\ \bibinfo {author} {\bibfnamefont {D.}~\bibnamefont {Loss}},\ }\href
  {\doibase 10.1103/PhysRevB.80.165119} {\bibfield  {journal} {\bibinfo
  {journal} {Phys. Rev. B}\ }\textbf {\bibinfo {volume} {80}},\ \bibinfo
  {pages} {165119} (\bibinfo {year} {2009}{\natexlab{a}})}\BibitemShut
  {NoStop}%
\bibitem [{\citenamefont {Braunecker}\ \emph
  {et~al.}(2009{\natexlab{b}})\citenamefont {Braunecker}, \citenamefont
  {Simon},\ and\ \citenamefont {Loss}}]{braunecker09b}%
  \BibitemOpen
  \bibfield  {author} {\bibinfo {author} {\bibfnamefont {B.}~\bibnamefont
  {Braunecker}}, \bibinfo {author} {\bibfnamefont {P.}~\bibnamefont {Simon}}, \
  and\ \bibinfo {author} {\bibfnamefont {D.}~\bibnamefont {Loss}},\ }\href
  {\doibase 10.1103/PhysRevLett.102.116403} {\bibfield  {journal} {\bibinfo
  {journal} {Phys. Rev. Lett.}\ }\textbf {\bibinfo {volume} {102}},\ \bibinfo
  {pages} {116403} (\bibinfo {year} {2009}{\natexlab{b}})}\BibitemShut
  {NoStop}%
\bibitem [{\citenamefont {Braunecker}\ \emph {et~al.}(2012)\citenamefont
  {Braunecker}, \citenamefont {Bena},\ and\ \citenamefont
  {Simon}}]{braunecker12}%
  \BibitemOpen
  \bibfield  {author} {\bibinfo {author} {\bibfnamefont {B.}~\bibnamefont
  {Braunecker}}, \bibinfo {author} {\bibfnamefont {C.}~\bibnamefont {Bena}}, \
  and\ \bibinfo {author} {\bibfnamefont {P.}~\bibnamefont {Simon}},\ }\href
  {\doibase 10.1103/PhysRevB.85.035136} {\bibfield  {journal} {\bibinfo
  {journal} {Phys. Rev. B}\ }\textbf {\bibinfo {volume} {85}},\ \bibinfo
  {pages} {035136} (\bibinfo {year} {2012})}\BibitemShut {NoStop}%
\bibitem [{\citenamefont {Scheller}\ \emph {et~al.}(2013)\citenamefont
  {Scheller}, \citenamefont {Liu}, \citenamefont {Barak}, \citenamefont
  {Yacoby}, \citenamefont {Pfeiffer}, \citenamefont {West},\ and\ \citenamefont
  {Zumb\"uhl}}]{scheller13}%
  \BibitemOpen
  \bibfield  {author} {\bibinfo {author} {\bibfnamefont {C.~P.}\ \bibnamefont
  {Scheller}}, \bibinfo {author} {\bibfnamefont {T.-M.}\ \bibnamefont {Liu}},
  \bibinfo {author} {\bibfnamefont {G.}~\bibnamefont {Barak}}, \bibinfo
  {author} {\bibfnamefont {A.}~\bibnamefont {Yacoby}}, \bibinfo {author}
  {\bibfnamefont {L.~N.}\ \bibnamefont {Pfeiffer}}, \bibinfo {author}
  {\bibfnamefont {K.~W.}\ \bibnamefont {West}}, \ and\ \bibinfo {author}
  {\bibfnamefont {D.~M.}\ \bibnamefont {Zumb\"uhl}},\ }\href
  {http://arxiv.org/abs/1306.1940} {\bibfield  {journal} {\bibinfo  {journal}
  {arXiv:1306.1940}\ } (\bibinfo {year} {2013})}\BibitemShut {NoStop}%
\bibitem [{\citenamefont {Gangadharaiah}\ \emph {et~al.}(2008)\citenamefont
  {Gangadharaiah}, \citenamefont {Sun},\ and\ \citenamefont
  {Starykh}}]{gangadharaiah08}%
  \BibitemOpen
  \bibfield  {author} {\bibinfo {author} {\bibfnamefont {S.}~\bibnamefont
  {Gangadharaiah}}, \bibinfo {author} {\bibfnamefont {J.}~\bibnamefont {Sun}},
  \ and\ \bibinfo {author} {\bibfnamefont {O.~A.}\ \bibnamefont {Starykh}},\
  }\href {\doibase 10.1103/PhysRevB.78.054436} {\bibfield  {journal} {\bibinfo
  {journal} {Phys. Rev. B}\ }\textbf {\bibinfo {volume} {78}},\ \bibinfo
  {pages} {054436} (\bibinfo {year} {2008})}\BibitemShut {NoStop}%
\bibitem [{\citenamefont {Stoudenmire}\ \emph {et~al.}(2011)\citenamefont
  {Stoudenmire}, \citenamefont {Alicea}, \citenamefont {Starykh},\ and\
  \citenamefont {Fisher}}]{stoudenmire11}%
  \BibitemOpen
  \bibfield  {author} {\bibinfo {author} {\bibfnamefont {E.~M.}\ \bibnamefont
  {Stoudenmire}}, \bibinfo {author} {\bibfnamefont {J.}~\bibnamefont {Alicea}},
  \bibinfo {author} {\bibfnamefont {O.~A.}\ \bibnamefont {Starykh}}, \ and\
  \bibinfo {author} {\bibfnamefont {M.~P.}\ \bibnamefont {Fisher}},\ }\href
  {\doibase 10.1103/PhysRevB.84.014503} {\bibfield  {journal} {\bibinfo
  {journal} {Phys. Rev. B}\ }\textbf {\bibinfo {volume} {84}},\ \bibinfo
  {pages} {014503} (\bibinfo {year} {2011})}\BibitemShut {NoStop}%
\bibitem [{\citenamefont {Meng}\ \emph {et~al.}(2013)\citenamefont {Meng},
  \citenamefont {Fritz}, \citenamefont {Schuricht},\ and\ \citenamefont
  {Loss}}]{meng13a}%
  \BibitemOpen
  \bibfield  {author} {\bibinfo {author} {\bibfnamefont {T.}~\bibnamefont
  {Meng}}, \bibinfo {author} {\bibfnamefont {L.}~\bibnamefont {Fritz}},
  \bibinfo {author} {\bibfnamefont {D.}~\bibnamefont {Schuricht}}, \ and\
  \bibinfo {author} {\bibfnamefont {D.}~\bibnamefont {Loss}},\ }\href
  {http://arxiv.org/abs/1308.3169} {\bibfield  {journal} {\bibinfo  {journal}
  {arXiv:1308.3169}\ } (\bibinfo {year} {2013})}\BibitemShut {NoStop}%
\bibitem [{\citenamefont {Ponomarenko}(1995)}]{ponomarenko95}%
  \BibitemOpen
  \bibfield  {author} {\bibinfo {author} {\bibfnamefont {V.~V.}\ \bibnamefont
  {Ponomarenko}},\ }\href {\doibase 10.1103/PhysRevB.52.R8666} {\bibfield
  {journal} {\bibinfo  {journal} {Phys. Rev. B}\ }\textbf {\bibinfo {volume}
  {52}},\ \bibinfo {pages} {R8666} (\bibinfo {year} {1995})}\BibitemShut
  {NoStop}%
\bibitem [{\citenamefont {Safi}\ and\ \citenamefont {Schulz}(1995)}]{safi95}%
  \BibitemOpen
  \bibfield  {author} {\bibinfo {author} {\bibfnamefont {I.}~\bibnamefont
  {Safi}}\ and\ \bibinfo {author} {\bibfnamefont {H.~J.}\ \bibnamefont
  {Schulz}},\ }\href {\doibase 10.1103/PhysRevB.52.R17040} {\bibfield
  {journal} {\bibinfo  {journal} {Phys. Rev. B}\ }\textbf {\bibinfo {volume}
  {52}},\ \bibinfo {pages} {R17040} (\bibinfo {year} {1995})}\BibitemShut
  {NoStop}%
\bibitem [{\citenamefont {Maslov}\ and\ \citenamefont
  {Stone}(1995)}]{maslov95}%
  \BibitemOpen
  \bibfield  {author} {\bibinfo {author} {\bibfnamefont {D.~L.}\ \bibnamefont
  {Maslov}}\ and\ \bibinfo {author} {\bibfnamefont {M.}~\bibnamefont {Stone}},\
  }\href {\doibase 10.1103/PhysRevB.52.R5539} {\bibfield  {journal} {\bibinfo
  {journal} {Phys. Rev. B}\ }\textbf {\bibinfo {volume} {52}},\ \bibinfo
  {pages} {R5539} (\bibinfo {year} {1995})}\BibitemShut {NoStop}%
\bibitem [{\citenamefont {Giamarchi}(2003)}]{giamarchi03}%
  \BibitemOpen
  \bibfield  {author} {\bibinfo {author} {\bibfnamefont {T.}~\bibnamefont
  {Giamarchi}},\ }\href@noop {} {\emph {\bibinfo {title} {Quantum Physics in
  One Dimension}}}\ (\bibinfo  {publisher} {Clarendon Press},\ \bibinfo
  {address} {Oxford},\ \bibinfo {year} {2003})\BibitemShut {NoStop}%
\bibitem [{\citenamefont {Khodas}\ \emph {et~al.}(2007)\citenamefont {Khodas},
  \citenamefont {Pustilnik}, \citenamefont {Kamenev},\ and\ \citenamefont
  {Glazman}}]{khodas07_2}%
  \BibitemOpen
  \bibfield  {author} {\bibinfo {author} {\bibfnamefont {M.}~\bibnamefont
  {Khodas}}, \bibinfo {author} {\bibfnamefont {M.}~\bibnamefont {Pustilnik}},
  \bibinfo {author} {\bibfnamefont {A.}~\bibnamefont {Kamenev}}, \ and\
  \bibinfo {author} {\bibfnamefont {L.~I.}\ \bibnamefont {Glazman}},\ }\href
  {\doibase 10.1103/PhysRevB.76.155402} {\bibfield  {journal} {\bibinfo
  {journal} {Phys. Rev. B}\ }\textbf {\bibinfo {volume} {76}},\ \bibinfo
  {pages} {155402} (\bibinfo {year} {2007})}\BibitemShut {NoStop}%
\bibitem [{\citenamefont {Barak}\ \emph {et~al.}(2010)\citenamefont {Barak},
  \citenamefont {Steinberg}, \citenamefont {Pfeiffer}, \citenamefont {West},
  \citenamefont {Glazman}, \citenamefont {von Oppen},\ and\ \citenamefont
  {Yacoby}}]{barak10}%
  \BibitemOpen
  \bibfield  {author} {\bibinfo {author} {\bibfnamefont {G.}~\bibnamefont
  {Barak}}, \bibinfo {author} {\bibfnamefont {H.}~\bibnamefont {Steinberg}},
  \bibinfo {author} {\bibfnamefont {L.~N.}\ \bibnamefont {Pfeiffer}}, \bibinfo
  {author} {\bibfnamefont {K.~W.}\ \bibnamefont {West}}, \bibinfo {author}
  {\bibfnamefont {L.}~\bibnamefont {Glazman}}, \bibinfo {author} {\bibfnamefont
  {F.}~\bibnamefont {von Oppen}}, \ and\ \bibinfo {author} {\bibfnamefont
  {A.}~\bibnamefont {Yacoby}},\ }\href {\doibase 10.1038/nphys1678} {\bibfield
  {journal} {\bibinfo  {journal} {Nature Physics}\ }\textbf {\bibinfo {volume}
  {6}},\ \bibinfo {pages} {489} (\bibinfo {year} {2010})}\BibitemShut {NoStop}%
\bibitem [{\citenamefont {Karzig}\ \emph {et~al.}(2010)\citenamefont {Karzig},
  \citenamefont {Glazman},\ and\ \citenamefont {von Oppen}}]{karzig10}%
  \BibitemOpen
  \bibfield  {author} {\bibinfo {author} {\bibfnamefont {T.}~\bibnamefont
  {Karzig}}, \bibinfo {author} {\bibfnamefont {L.~I.}\ \bibnamefont {Glazman}},
  \ and\ \bibinfo {author} {\bibfnamefont {F.}~\bibnamefont {von Oppen}},\
  }\href {\doibase 10.1103/PhysRevLett.105.226407} {\bibfield  {journal}
  {\bibinfo  {journal} {Phys. Rev. Lett.}\ }\textbf {\bibinfo {volume} {105}},\
  \bibinfo {pages} {226407} (\bibinfo {year} {2010})}\BibitemShut {NoStop}%
\bibitem [{\citenamefont {Schmidt}\ \emph {et~al.}(2010)\citenamefont
  {Schmidt}, \citenamefont {Imambekov},\ and\ \citenamefont
  {Glazman}}]{schmidt10_2}%
  \BibitemOpen
  \bibfield  {author} {\bibinfo {author} {\bibfnamefont {T.~L.}\ \bibnamefont
  {Schmidt}}, \bibinfo {author} {\bibfnamefont {A.}~\bibnamefont {Imambekov}},
  \ and\ \bibinfo {author} {\bibfnamefont {L.~I.}\ \bibnamefont {Glazman}},\
  }\href {\doibase 10.1103/PhysRevB.82.245104} {\bibfield  {journal} {\bibinfo
  {journal} {Phys. Rev. B}\ }\textbf {\bibinfo {volume} {82}},\ \bibinfo
  {pages} {245104} (\bibinfo {year} {2010})}\BibitemShut {NoStop}%
\bibitem [{\citenamefont {Imambekov}\ \emph {et~al.}(2012)\citenamefont
  {Imambekov}, \citenamefont {Schmidt},\ and\ \citenamefont
  {Glazman}}]{imambekov12}%
  \BibitemOpen
  \bibfield  {author} {\bibinfo {author} {\bibfnamefont {A.}~\bibnamefont
  {Imambekov}}, \bibinfo {author} {\bibfnamefont {T.~L.}\ \bibnamefont
  {Schmidt}}, \ and\ \bibinfo {author} {\bibfnamefont {L.~I.}\ \bibnamefont
  {Glazman}},\ }\href {\doibase 10.1103/RevModPhys.84.1253} {\bibfield
  {journal} {\bibinfo  {journal} {Rev. Mod. Phys.}\ }\textbf {\bibinfo {volume}
  {84}},\ \bibinfo {pages} {1253} (\bibinfo {year} {2012})}\BibitemShut
  {NoStop}%
\bibitem [{\citenamefont {Ristivojevic}\ and\ \citenamefont
  {Matveev}(2013)}]{ristivojevic13}%
  \BibitemOpen
  \bibfield  {author} {\bibinfo {author} {\bibfnamefont {Z.}~\bibnamefont
  {Ristivojevic}}\ and\ \bibinfo {author} {\bibfnamefont {K.~A.}\ \bibnamefont
  {Matveev}},\ }\href {\doibase 10.1103/PhysRevB.87.165108} {\bibfield
  {journal} {\bibinfo  {journal} {Phys. Rev. B}\ }\textbf {\bibinfo {volume}
  {87}},\ \bibinfo {pages} {165108} (\bibinfo {year} {2013})}\BibitemShut
  {NoStop}%
\bibitem [{\citenamefont {Lunde}\ \emph {et~al.}(2007)\citenamefont {Lunde},
  \citenamefont {Flensberg},\ and\ \citenamefont {Glazman}}]{lunde07}%
  \BibitemOpen
  \bibfield  {author} {\bibinfo {author} {\bibfnamefont {A.~M.}\ \bibnamefont
  {Lunde}}, \bibinfo {author} {\bibfnamefont {K.}~\bibnamefont {Flensberg}}, \
  and\ \bibinfo {author} {\bibfnamefont {L.~I.}\ \bibnamefont {Glazman}},\
  }\href {\doibase 10.1103/PhysRevB.75.245418} {\bibfield  {journal} {\bibinfo
  {journal} {Phys. Rev. B}\ }\textbf {\bibinfo {volume} {75}},\ \bibinfo
  {pages} {245418} (\bibinfo {year} {2007})}\BibitemShut {NoStop}%
\bibitem [{\citenamefont {Rech}\ \emph {et~al.}(2009)\citenamefont {Rech},
  \citenamefont {Micklitz},\ and\ \citenamefont {Matveev}}]{rech09}%
  \BibitemOpen
  \bibfield  {author} {\bibinfo {author} {\bibfnamefont {J.}~\bibnamefont
  {Rech}}, \bibinfo {author} {\bibfnamefont {T.}~\bibnamefont {Micklitz}}, \
  and\ \bibinfo {author} {\bibfnamefont {K.~A.}\ \bibnamefont {Matveev}},\
  }\href {\doibase 10.1103/PhysRevLett.102.116402} {\bibfield  {journal}
  {\bibinfo  {journal} {Phys. Rev. Lett.}\ }\textbf {\bibinfo {volume} {102}},\
  \bibinfo {pages} {116402} (\bibinfo {year} {2009})}\BibitemShut {NoStop}%
\bibitem [{\citenamefont {Micklitz}\ \emph {et~al.}(2010)\citenamefont
  {Micklitz}, \citenamefont {Rech},\ and\ \citenamefont
  {Matveev}}]{micklitz10}%
  \BibitemOpen
  \bibfield  {author} {\bibinfo {author} {\bibfnamefont {T.}~\bibnamefont
  {Micklitz}}, \bibinfo {author} {\bibfnamefont {J.}~\bibnamefont {Rech}}, \
  and\ \bibinfo {author} {\bibfnamefont {K.~A.}\ \bibnamefont {Matveev}},\
  }\href {\doibase 10.1103/PhysRevB.81.115313} {\bibfield  {journal} {\bibinfo
  {journal} {Phys. Rev. B}\ }\textbf {\bibinfo {volume} {81}},\ \bibinfo
  {pages} {115313} (\bibinfo {year} {2010})}\BibitemShut {NoStop}%
\bibitem [{\citenamefont {Micklitz}\ and\ \citenamefont
  {Levchenko}(2011)}]{micklitz11}%
  \BibitemOpen
  \bibfield  {author} {\bibinfo {author} {\bibfnamefont {T.}~\bibnamefont
  {Micklitz}}\ and\ \bibinfo {author} {\bibfnamefont {A.}~\bibnamefont
  {Levchenko}},\ }\href {\doibase 10.1103/PhysRevLett.106.196402} {\bibfield
  {journal} {\bibinfo  {journal} {Phys. Rev. Lett.}\ }\textbf {\bibinfo
  {volume} {106}},\ \bibinfo {pages} {196402} (\bibinfo {year}
  {2011})}\BibitemShut {NoStop}%
\bibitem [{\citenamefont {Matveev}\ and\ \citenamefont
  {Andreev}(2011)}]{matveev11}%
  \BibitemOpen
  \bibfield  {author} {\bibinfo {author} {\bibfnamefont {K.~A.}\ \bibnamefont
  {Matveev}}\ and\ \bibinfo {author} {\bibfnamefont {A.~V.}\ \bibnamefont
  {Andreev}},\ }\href {\doibase 10.1103/PhysRevLett.107.056402} {\bibfield
  {journal} {\bibinfo  {journal} {Phys. Rev. Lett.}\ }\textbf {\bibinfo
  {volume} {107}},\ \bibinfo {pages} {056402} (\bibinfo {year}
  {2011})}\BibitemShut {NoStop}%
\bibitem [{\citenamefont {Schuricht}(2012)}]{schuricht12}%
  \BibitemOpen
  \bibfield  {author} {\bibinfo {author} {\bibfnamefont {D.}~\bibnamefont
  {Schuricht}},\ }\href {\doibase 10.1103/PhysRevB.85.121101} {\bibfield
  {journal} {\bibinfo  {journal} {Phys. Rev. B}\ }\textbf {\bibinfo {volume}
  {85}},\ \bibinfo {pages} {121101(R)} (\bibinfo {year} {2012})}\BibitemShut
  {NoStop}%
\bibitem [{\citenamefont {Liu}\ \emph {et~al.}(2012)\citenamefont {Liu},
  \citenamefont {Potter}, \citenamefont {Law},\ and\ \citenamefont
  {Lee}}]{liu12a}%
  \BibitemOpen
  \bibfield  {author} {\bibinfo {author} {\bibfnamefont {J.}~\bibnamefont
  {Liu}}, \bibinfo {author} {\bibfnamefont {A.~C.}\ \bibnamefont {Potter}},
  \bibinfo {author} {\bibfnamefont {K.~T.}\ \bibnamefont {Law}}, \ and\
  \bibinfo {author} {\bibfnamefont {P.~A.}\ \bibnamefont {Lee}},\ }\href
  {\doibase 10.1103/PhysRevLett.109.267002} {\bibfield  {journal} {\bibinfo
  {journal} {Phys. Rev. Lett.}\ }\textbf {\bibinfo {volume} {109}},\ \bibinfo
  {pages} {267002} (\bibinfo {year} {2012})}\BibitemShut {NoStop}%
\bibitem [{\citenamefont {Bagrets}\ and\ \citenamefont
  {Altland}(2012)}]{bagrets12}%
  \BibitemOpen
  \bibfield  {author} {\bibinfo {author} {\bibfnamefont {D.}~\bibnamefont
  {Bagrets}}\ and\ \bibinfo {author} {\bibfnamefont {A.}~\bibnamefont
  {Altland}},\ }\href {\doibase 10.1103/PhysRevLett.109.227005} {\bibfield
  {journal} {\bibinfo  {journal} {Phys. Rev. Lett.}\ }\textbf {\bibinfo
  {volume} {109}},\ \bibinfo {pages} {227005} (\bibinfo {year}
  {2012})}\BibitemShut {NoStop}%
\bibitem [{\citenamefont {Bruus}\ and\ \citenamefont
  {Flensberg}(2004)}]{bruus04}%
  \BibitemOpen
  \bibfield  {author} {\bibinfo {author} {\bibfnamefont {H.}~\bibnamefont
  {Bruus}}\ and\ \bibinfo {author} {\bibfnamefont {K.}~\bibnamefont
  {Flensberg}},\ }\href@noop {} {\emph {\bibinfo {title} {Many-body quantum
  theory in condensed matter physics}}}\ (\bibinfo  {publisher} {Oxford
  University Press},\ \bibinfo {year} {2004})\BibitemShut {NoStop}%
\bibitem [{\citenamefont {Abrikosov}(1972)}]{abrikosov_book}%
  \BibitemOpen
  \bibfield  {author} {\bibinfo {author} {\bibfnamefont {A.~A.}\ \bibnamefont
  {Abrikosov}},\ }\href@noop {} {\emph {\bibinfo {title} {Introduction to the
  theory of normal metals}}}\ (\bibinfo  {publisher} {Academic Press},\
  \bibinfo {address} {New York and London},\ \bibinfo {year}
  {1972})\BibitemShut {NoStop}%
\bibitem [{\citenamefont {Landauer}(1987)}]{landauer87}%
  \BibitemOpen
  \bibfield  {author} {\bibinfo {author} {\bibfnamefont {R.}~\bibnamefont
  {Landauer}},\ }\href {\doibase 10.1007/BF01304229} {\bibfield  {journal}
  {\bibinfo  {journal} {Z. Phys. B}\ }\textbf {\bibinfo {volume} {68}},\
  \bibinfo {pages} {217} (\bibinfo {year} {1987})}\BibitemShut {NoStop}%
\bibitem [{\citenamefont {Glazman}\ \emph {et~al.}(1988)\citenamefont
  {Glazman}, \citenamefont {Lesovik}, \citenamefont {Khmel'nitskii},\ and\
  \citenamefont {Shekhter}}]{glazman88}%
  \BibitemOpen
  \bibfield  {author} {\bibinfo {author} {\bibfnamefont {L.~I.}\ \bibnamefont
  {Glazman}}, \bibinfo {author} {\bibfnamefont {G.~B.}\ \bibnamefont
  {Lesovik}}, \bibinfo {author} {\bibfnamefont {D.~E.}\ \bibnamefont
  {Khmel'nitskii}}, \ and\ \bibinfo {author} {\bibfnamefont {R.~I.}\
  \bibnamefont {Shekhter}},\ }\href
  {http://www.jetpletters.ac.ru/ps/1104/article_16696.shtml} {\bibfield
  {journal} {\bibinfo  {journal} {JETP Lett.}\ }\textbf {\bibinfo {volume}
  {48}},\ \bibinfo {pages} {238} (\bibinfo {year} {1988})}\BibitemShut
  {NoStop}%
\bibitem [{\citenamefont {Lunde}\ \emph {et~al.}(2006)\citenamefont {Lunde},
  \citenamefont {Flensberg},\ and\ \citenamefont {Glazman}}]{lunde06}%
  \BibitemOpen
  \bibfield  {author} {\bibinfo {author} {\bibfnamefont {A.~M.}\ \bibnamefont
  {Lunde}}, \bibinfo {author} {\bibfnamefont {K.}~\bibnamefont {Flensberg}}, \
  and\ \bibinfo {author} {\bibfnamefont {L.~I.}\ \bibnamefont {Glazman}},\
  }\href {\doibase 10.1103/PhysRevLett.97.256802} {\bibfield  {journal}
  {\bibinfo  {journal} {Phys. Rev. Lett.}\ }\textbf {\bibinfo {volume} {97}},\
  \bibinfo {pages} {256802} (\bibinfo {year} {2006})}\BibitemShut {NoStop}%
\bibitem [{\citenamefont {Rainis}\ and\ \citenamefont {Loss}(2012)}]{rainis12}%
  \BibitemOpen
  \bibfield  {author} {\bibinfo {author} {\bibfnamefont {D.}~\bibnamefont
  {Rainis}}\ and\ \bibinfo {author} {\bibfnamefont {D.}~\bibnamefont {Loss}},\
  }\href {\doibase 10.1103/PhysRevB.85.174533} {\bibfield  {journal} {\bibinfo
  {journal} {Phys. Rev. B}\ }\textbf {\bibinfo {volume} {85}},\ \bibinfo
  {pages} {174533} (\bibinfo {year} {2012})}\BibitemShut {NoStop}%
\bibitem [{\citenamefont {Yang}(1967)}]{yang67}%
  \BibitemOpen
  \bibfield  {author} {\bibinfo {author} {\bibfnamefont {C.~N.}\ \bibnamefont
  {Yang}},\ }\href {\doibase 10.1103/PhysRevLett.19.1312} {\bibfield  {journal}
  {\bibinfo  {journal} {Phys. Rev. Lett.}\ }\textbf {\bibinfo {volume} {19}},\
  \bibinfo {pages} {1312} (\bibinfo {year} {1967})}\BibitemShut {NoStop}%
\bibitem [{\citenamefont {Gaudin}(1967)}]{gaudin67}%
  \BibitemOpen
  \bibfield  {author} {\bibinfo {author} {\bibfnamefont {M.}~\bibnamefont
  {Gaudin}},\ }\href {\doibase 10.1016/0375-9601(67)90193-4} {\bibfield
  {journal} {\bibinfo  {journal} {Phys. Lett. A}\ }\textbf {\bibinfo {volume}
  {24}},\ \bibinfo {pages} {55} (\bibinfo {year} {1967})}\BibitemShut {NoStop}%
\bibitem [{\citenamefont {Sutherland}(2004)}]{sutherland04}%
  \BibitemOpen
  \bibfield  {author} {\bibinfo {author} {\bibfnamefont {B.}~\bibnamefont
  {Sutherland}},\ }\href@noop {} {\emph {\bibinfo {title} {Beautiful Models: 70
  Years Of Exactly Solved Quantum Many-Body Problems}}}\ (\bibinfo  {publisher}
  {World Scientific},\ \bibinfo {year} {2004})\BibitemShut {NoStop}%
\end{thebibliography}%

\end{document}